\documentclass[usenatbib,useAMS,usegraphicx]{mn2e}

\title[Formation of the Radio Profile Components of the Crab Pulsar]
{Formation of the Radio Profile Components of the Crab Pulsar}

\author[S. A. Petrova]{S. A. Petrova
\thanks{E-mail: petrova@ri.kharkov.ua}\\
Institute of Radio Astronomy, NAS of Ukraine, 4, Chervonopraporna
Str., 61002 Kharkov, Ukraine}

\begin{document}

\date{Received\dots}

\pagerange{\pageref{firstpage}--\pageref{lastpage}} \pubyear{2009}

\maketitle

\protect\label{firstpage}

\begin{abstract}
The induced Compton scattering of radio emission off the particles of the ultrarelativistic electron-positron plasma in the open field line tube of a pulsar is considered. We examine the scattering of a bright narrow radio beam into the background over a wide solid angle and specifically study the scattering in the transverse regime, which holds in a moderately strong magnetic field and gives rise to the scattered component nearly antiparallel to the streaming velocity of the scattering particles. Making use of the angular distribution of the scattered intensity and taking into account the effect of rotational aberration in the scattering region, we simulate the profiles of the backscattered components as applied to the Crab pulsar. It is suggested that the interpulse (IP), the high-frequency interpulse (IP') and the pair of the so-called high-frequency components (HFC1 and HFC2) result from the backward scattering of the main pulse (MP), precursor (PR) and the low-frequency component (LFC), respectively. The components of the high-frequency profiles, the IP' and HFCs, are interpreted for the first time. The HFC1 and HFC2 are argued to be a single component split by the rotational aberration close to the light cylinder. It is demonstrated that the observed spectral and polarization properties of the profile components of the Crab pulsar as well as the giant pulse phenomenon outside of the MP can be explained in terms of our model.
\end{abstract}


\section{Introduction}

The radio emission pattern of the Crab pulsar is unique in its complexity. There are seven emission components located at different longitudinal regions throughout the whole pulse period and characterized by distinct spectra and polarization \citep[e.g.][]{mh96,mh99}. At the lowest frequencies, the radio profile of the Crab pulsar consists of three components: the main pulse (MP), the precursor (PR) $\sim 15^\circ$ ahead, and the interpulse (IP) $\sim 150^\circ$ behind the MP. Such a structure is also characteristic of the profiles of some other pulsars (e.g., PSR B1055-52, \citealt{m76} and PSR B1822-09,
\citealt*{f81}). However in the Crab pulsar the components have exceptionally steep spectra, with the power law indices $\alpha_\mathrm{PR}=-5$, $\alpha_\mathrm{IP}=4.1$ and $\alpha_\mathrm{MP}=-3$, and vanish at the frequencies $\sim 0.6$, $3$ and $8$ GHz, respectively. In the range 1-3 GHz, the Crab profile contains the so-called low-frequency component (LFC) $\sim 36^\circ$ in advance of the MP. At still higher frequencies, there appear the high-frequency interpulse (IP$^\prime$) $\sim 10^\circ$ earlier in pulse longitude than its low-frequency analogue (the IP) and a pair of the so-called high-frequency components (the HFC1 and HFC2), which lag the IP$^\prime$ by $\sim 70^\circ$ and $130^\circ$, respectively. The IP$^\prime$, HFC1 and HFC2 have flat spectra and dominate the pulse profile at the frequencies $\ga 8$ GHz.

The radio polarization studies of the Crab pulsar \citep*{chr70,m71a,mht72,mh99,kjw04} have revealed moderate linear polarization of the MP, IP and LFC ($\sim 15\%$, $25\%$ and $40\%$, respectively), high polarization of the HFC1,2 and complete linear polarization of the PR and IP$^\prime$. In the MP, PR and IP, the position angle of linear polarization is nearly constant across the components and is nearly the same for all of them. In the rest of the components and between them, the position angle noticeably changes with pulse longitude. In the LFC and IP$^\prime$, the characteristic values of the position angle differ from that of the MP by $\sim 30^\circ$ and $90^\circ$, respectively, whereas in the HFC1 and HFC2 the position angle is shifted by $\sim 90^\circ$ from that of the LFC. A straightforward interpretation of the position angle swing in terms of the primordial rotating vector model \citep{rc69} faces difficulties \citep{mh99,kjw04}. 

In spite of substantial distinctions in the spectra and polarization, most of the components exhibit similar fluctuation properties. The Crab pulsar is known as a source of giant pulses -- the occasional strong radio pulses with intensities ranging from a few dozen to a few thousand times the average \citep[e.g.,][]{l95}. The giant pulse activity is characteristic of the MP, IP and IP$^\prime$ \citep{s99,c04,k08}, and it has also been found in the HFC1 and HFC2 \citep{j05}. In these components, the giant pulses may be present simultaneously, but the rate of their occurrence is substantially different \citep{s99,c04,popov06,k08}. Furthermore, the temporal and frequency structure of the giant emission in the MP and IP$^\prime$ is essentially distinct \citep{eh07}. The giant MPs present one or several broadband microbursts,which consist of narrowband ($\delta\nu/\nu\sim 0.1$) shots of a nanosecond duration. The giant IP$^\prime$s consist of the proportionally spaced narrow emission bands of microsecond lengths organized into several band sets.

The radio profile components of pulsars outside of the MP used to be interpreted in terms of several geometrical models \citep[e.g.,][]{hc81}, which assumed a  peculiar geometry of the pulsar (either orthogonality or alignment of its magnetic and rotation axes) and placed the emission regions of the components at different sites in the magnetosphere. Recently this approach was complemented by the concept of inward emission \citep*{d05}: although pulsar radiation is undoubtedly emitted along the magnetic lines, it may well be directed backwards, toward the neutron star. However the complicated radio emission pattern of the Crab pulsar remains beyond the framework of any plausible geometrical model. Moreover, the comprehensive spectral, polarization and fluctuation data testify against the independent generation of the components and highlight the question of their physical nature.

Recently we have proposed a physical mechanism of the PR and IP components in radio pulsars \citep{p08a,p08b}. It is based on the propagation effects in the flow of pair plasma inside the open field line tube of a pulsar. The PR and IP are suggested to result from the induced scattering of the MP radiation into background. The scattered radiation strongly concentrates in the direction corresponding to the maximum scattering probability and forms separate components in the pulse profile. In the regime of a superstrong magnetic field, the scattered radiation is directed approximately along the field, and, taking into account the effect of rotational aberration, one can recognize it as a PR component. In a moderately strong magnetic field, the radiation is predominantly scattered backwards -- in the direction antiparallel to the velocity of the scattering particles, which stream along the magnetic lines, -- and the scattered component can be identified with the IP.

In application to the Crab pulsar, we have generalized the problem of induced scattering by taking into account the helical motion of the scattering particles \citep{p08c}. In the outer magnetosphere, the particles acquire gyration energies as a result of resonant absorption of radio emission. They arrive at relativistic gyration at the very bottom of the resonance region, where the highest radio frequencies are absorbed. At higher altitudes, the character of the radio wave scattering changes essentially: now the waves can be scattered between different harmonics of the particle gyroferquency. It has been shown that the induced scattering between the states well below the resonance may still account for the PR component in the radio profile of the Crab pulsar, whereas the scattering from the first harmonic to the zeroth one may be responsible for the LFC \citep{p08c}.

As the radiation of the PR and LFC appears well below the resonance, further on its way in the magnetosphere it is subject to the induced scattering in a moderately strong magnetic field. Similarly to the MP scattering, it may give rise to the additional backward components in the pulse profile. A preliminary analysis of the Crab polarization data suggests that the IP$^\prime$ results from the backward scattering of the PR, whereas the pair of HFCs is a consequence of the backscattering of the LFC.

It should be noted, however, that the simplified scattering geometry derived in \citet{p08b} cannot be directly extended to the case of the Crab pulsar: neither of the backscattered components, including the IP, can be properly located in the pulse profile. Given that the backscattered radiation is directed along the magnetic lines and suffers rotational aberration, the scattered component should lag the original one by more than $180^\circ$ in pulse longitude \citep[see also][]{d05}. At the same time, the IP of the Crab pulsar is separated from the MP by only $\sim 150^\circ$, the distance between the IP$^\prime$ and PR is almost the same and the separation of the HFC1,2 from the LFC, being larger than $180^\circ$, is still small to account for the scattering above the LFC origin.

In the present paper, we overcome this seeming difficulty by means of a more accurate examination of the component formation. The paper presents a detailed study of the intensity transfer in the course of induced scattering of pulsar radiation into the background. On this basis we elaborate our theory of the pulsar radio profile components outside of the MP, extending it to the case of the Crab pulsar, with an especial emphasis on the formation of its high-frequency components (the IP$^\prime$, HFC1 and HFC2). In Sect.~\ref{s2} we solve the problem of induced scattering of a narrow radio beam into a wide solid angle in the presence of a moderately strong magnetic field. With the angular distribution of the scattered intensity in hand, in Sect.~\ref{s3} we simulate the profiles of the high-frequency components of the Crab pulsar, taking into account the effect of rotational aberration. The spectral, polarization and fluctuation properties of the scattered components are confronted with the observational evidence in Sect.~\ref{s4}. The results are briefly summarized in Sect.~\ref{s5}.

\section{Induced scattering into a wide solid angle}
\protect\label{s2}

\subsection{Statement of the problem \protect\label{s2.1}}

Pulsar magnetosphere contains the ultrarelativistic electron-positron plasma, which streams along the open magnetic lines. The radio emission originates deep in the magnetosphere and subsequently propagates through the plasma flow. The radio wave scattering off the plasma particles is one of the significant propagation effects in pulsars. As the brightness temperatures of pulsar radio emission are extremely high, the induced scattering strongly dominates the spontaneous one \citep[e.g.,][]{wr78}.

The external magnetic field may affect the scattering process considerably provided that the frequency of incident waves in the rest frame of the scattering particles is less than the particle gyrofrequency, $\omega^\prime\equiv\omega\gamma (1-\beta\cos\theta)\ll\omega_G\equiv eB/mc$ (here $\omega$ is the wave frequency in the laboratory frame, $\beta$ is the particle velocity in units of $c$, $\gamma$ is the Lorentz-factor of the particle motion, $\gamma\equiv(1-\beta^2)^{-1/2}$, $\theta$ is the angle of incidence of the waves). The radio waves typically meet the cyclotron resonance condition, $\omega^\prime=\omega_G$, in the outer magnetosphere, and therefore the pulsar radio emission is subject to both the magnetized and non-magnetized scattering. The non-magnetized induced scattering was considered in \citet{wr78,w82,hm95,lp96,l08}, and the magnetized scattering was addressed in \citet{bs76,lp96,p04a,p04b,p08a,p08b}.

The presence of a strong magnetic field can modify the scattering process variously \citep[e.g.,][]{bs76}. In an extremely strong field, the perturbed motion of a particle in the fields of the incident wave is confined to the magnetic line. Due to such a character of the particle motion the scattering process has a number of peculiarities and is usually called the longitudinal scattering. In this case, only the waves of the ordinary polarization, with the electric vector in the plane of the ambient magnetic field, suffer the scattering, and the scattered waves also have the same polarization. Besides that, the waves are predominantly scattered forwards, at $\sim 1/\gamma$ to the magnetic field direction. In a moderately strong magnetic field, the components of the particle's perturbed motion perpendicular to the magnetic line affect the characteristics of the scattered radiation significantly. Both the ordinary and extraordinary waves participate in the scattering process, and the induced scattering is most probable in the backward direction. This is a so-called transverse scattering. The induced scattering switches between the regimes on condition that $\gamma^2\omega^{\prime^4}/\omega_G^4=1$ \citep{p08b}, both regimes being typical of the pulsar magnetosphere.

Pulsar radio emission is known to be highly directional. At any point of the pulsar emission cone it is concentrated into a narrow beam of the opening angle $\sim 1/\gamma$. Therefore the induced scattering out of the beam dominates that inside the beam \citep{lp96}. The former process may only hold if initially there is some background radiation outside of the beam. This background radiation may result from the spontaneous scattering out of the beam. Although the background intensity is extremely small as compared to the beam intensity, $I_{\mathrm{bg}}/I_0=10^{-8}-10^{-12}$, it may still trigger an efficient induced scattering of the beam radiation into the background.

As soon as the induced scattering starts, the intensity of the background grows exponentially. It is clear that a substantial growth may occur only for the states with the photon orientations $\theta_1$ close to that corresponding to the maximum scattering probability, $\theta_1^{\mathrm{max}}$. In the rest frame of the scattering particles, the frequencies of the incident and scattered photons are almost the same, $\omega^\prime\approx\omega_1^\prime$. Hence, in the laboratory frame the photon states involved in the scattering satisfy the condition
\begin{equation}
\omega\gamma (1-\beta\cos\theta)=\omega_1\gamma(1-\beta\cos\theta_1).
\label{eq1}
\end{equation}
In our previous papers, we have considered the induced scattering between the two photon states satisfying equation~(\ref{eq1}), one of which represents the photon beam of a given frequency $\omega$ and orientation $\theta$ ($1/\gamma\ll\theta\la 1$) and another one the state of the most probable scattering, $\theta_1=\theta_1^{\mathrm{max}}$ \citep{p08a,p08b,p08c}. Then the broadband picture of the scattering has been reconstructed from the two-state solutions at different $\omega$.

It should be kept in mind, however, that the scattering efficiency depends not only on the orientation of the scattered photons, but also on the incident intensity, which is a function of frequency. For a given frequency of the background photons, $\omega_1$, different photon orientations $\theta_1$ correspond to different feeding frequencies $\omega$ (cf. Eq.~(\ref{eq1})) and, consequently, to different original intensities of the radio beam. If, say, $\theta_1^{\mathrm{max}}=\pi$, smaller $\theta_1$ imply smaller $\omega$ and larger $I_0$, so that the scattering into these directions may be even more efficient. Thus, the location of the scattered components in the pulse profile can be affected by the radio beam spectrum.

In the present paper, we take into account this effect and examine the induced scattering of a narrow radio beam of a fixed frequency into all conceivable states of the background, which satisfy equation~(\ref{eq1}). As the beam photons of different frequencies are scattered independently, the extension of our results to the broadband case is straightforward. The main motivation of our study is the necessity of modeling the profiles of the scattered components as applied to the Crab pulsar, which is known for its extremely steep radio spectrum. We consider the transverse scattering, which gives rise to the backward components. As for the longitudinal scattering, its efficiency is a much stronger function of the orientation of the scattered photons, and the radio beam spectrum does not affect the profile location of the scattered components considerably.

\subsection{Intensity transfer \protect\label{s2.2}}

Let us consider the transverse induced scattering in the frame corotating with the neutron star. The evolution of the photon occupation number $n$ in the course of the scattering is described by the kinetic equation (1) in \citet{p08b}. One can rewrite it in terms of intensity, $i_\nu=2h\nu^3n/c^2$, and integrate over the Lorentz-factor of the scattering particles, noticing that a detailed form of the particle distribution function does not play a crucial role and taking the monoenergetic distribution with some characteristic Lorentz-factor. The radio beam intensity is assumed to have the delta-functional angular distribution, and it is convenient to introduce the spectral intensity of the beam integrated over the solid angle, $I_\nu=\int i_{\nu_\mathrm{beam}}\mathrm{d}\Omega$. Pulsar radio emission is generally a mixture of the two types of waves: the ordinary ones polarized in the plane of the ambient magnetic field (the A-polarization) and the extraordinary ones polarized perpendicularly to that plane (the B-polarization). For the sake of simplicity we assume that the radio beam consists of the waves of one polarization and interacts with both polarization states of the background. Then the intensity transfer between the beam and the background can be presented as follows:
\[
\frac{\mathrm{d}I^i}{\mathrm{d}x}=-I^i\sum\limits_{j=A,B}\int i^j(\theta_1,\phi_1)(\cos\theta-\cos\theta_1)g^{ij}(\theta_1,\phi_1)\mathrm{d}\Omega_1,
\]
\[
\frac{\mathrm{d}i^j(\theta_{1_\alpha},\phi_{1_\alpha})}{\mathrm{d}x}=i^j(\theta_{1_\alpha},\phi_{1_\alpha})I^i(\cos\theta-\cos\theta_{1_\alpha})g^{ij}(\theta_{1_\alpha},\phi_{1_\alpha}),
\]
\begin{equation}
\frac{\mathrm{d}i^j(\theta_{1_\beta},\phi_{1_\beta})}{\mathrm{d}x}=i^j(\theta_{1_\beta},\phi_{1_\beta})I^i(\cos\theta-\cos\theta_{1_\beta})g^{ij}(\theta_{1_\beta},\phi_{1_\beta}),
\label{eq2}
\end{equation}
\[
\dots\dots\dots\dots\dots\dots\dots\dots\dots\dots\dots\dots\dots\dots\dots\dots\dots\dots
\]
This infinite set contains the equations for the background photons of any conceivable orientations $(\theta_{1_l},\phi_{1_l})$, $l=\alpha,\beta,\dots$. Here the intensities of the beam and the background are normalized to the original radio beam intensity, $I\equiv I_\nu/I_\nu^{(0)}$ and $i\equiv i_{\nu_\mathrm{bg}}/I_\nu^{(0)}$, $\mathrm{d}x\equiv I_\nu^{(0)}a\mathrm{d}r$,
\begin{equation}
a=\frac{4n_er_e^2\beta}{m\gamma^3\nu^2\theta^4}\frac{\nu^{\prime^4}}{\nu_G^4},
\label{eq3}
\end{equation}
$n_e$ is the number density of the scattering particles, $r_e$ is the classical electron radius, $\theta$ is the beam tilt to the magnetic field, $1/\gamma\ll\theta\la 1$, $\nu^\prime\equiv\nu\gamma(1-\beta\cos\theta)$, $\nu_G\equiv\omega_G/2\pi$, $r$ is the coordinate along the beam trajectory in the scattering region, the subscripts of the factor $g^{ij}$ stand for the polarization states of the incident and scattered photons,
\[
g^{AA}=(1+\sin^2\Delta\phi)\frac{(1-\eta\gamma^2)^2}{\beta^2\gamma^4\eta^2}
\frac{(1-\eta_1\gamma^2)^2}{\beta^2\gamma^4\eta_1^2}\]
\[-\frac{\sin
2\theta\sin 2\theta_1\cos\Delta\phi }{2},
\]
\[
g^{AB}=(1+\cos^2\Delta\phi)\frac{(1-\eta\gamma^2)^2}{\beta^2\gamma^4\eta^2},
\]
\[
g^{BA}=(1+\cos^2\Delta\phi)\frac{(1-\eta_1\gamma^2)^2}{\beta^2\gamma^4\eta_1^2},
\]
\begin{equation}
g^{BB}=1+\sin^2\Delta\phi ,
\label{eq4}
\end{equation}
$\Delta\phi\equiv\phi_1-\phi$ is the azimuthal coordinate of the background photons with respect to that of the beam, $\eta\equiv 1-\beta\cos\theta$ and $\eta_1\equiv 1-\beta\cos\theta_1$. In the set of equations (\ref{eq2}) it is taken into account that initially the background intensity is very small as compared to the beam intensity and therefore the terms describing the induced scattering between the states of the background photons are omitted. One can see that the total intensity of the beam and the background is conserved,
\begin{equation}
I^i+\sum\limits_j\int i^j\mathrm{d}\Omega_1\equiv C\approx 1.
\label{eq5}
\end{equation}
Furthermore, from the last two equations of the set (\ref{eq2}) one can find that for the two arbitrary orientations of the background photons, $(\theta_{1_\alpha},\phi_{1_\alpha})$ and $(\theta_{1_\beta},\phi_{1_\beta})$, the intensities are related as
\begin{equation}
\left [\frac{i^j(\theta_{1_\alpha},\phi_{1_\alpha})}{i_0^j(\theta_{1_\alpha},\phi_{1_\alpha})}\right ]^\frac{1}{(\cos\theta-\cos\theta_{1_\alpha})}\equiv \left [\frac{i^j(\theta_{1_\beta},\phi_{1_\beta})}{i_0^j(\theta_{1_\beta},\phi_{1_\beta})}\right ]^\frac{1}{(\cos\theta-\cos\theta_{1_\beta})}.
\label{eq6}
\end{equation}
Making the substitution
\begin{equation}
i^j(\theta_{1_l},\phi_{1_l})=i_0^j(\theta_{1_l},\phi_{1_l})\exp(y^ig^{ij}(\cos\theta-\cos\theta_{1_l}))
\label{eq7}
\end{equation}
and using equations (\ref{eq5})-(\ref{eq6}) in any of the equations (\ref{eq2}) for the background intensity, we obtain
\begin{equation}
\frac{\mathrm{d}y^i}{\mathrm{d}x}=1-\sum\limits_{j=A,B}\int i_0^j(\theta_1,\phi_1)\mathrm{e}^{y^ig^{ij}(\cos\theta-\cos\theta_1)}\mathrm{d}\Omega_1.
\label{eq8}
\end{equation}
The solution $y^i(x)$ completely determines the intensity evolution in the course of the scattering. For any point of the scattering region, the beam intensity reads
\begin{equation}
I^i=\frac{\mathrm{d}y^i}{\mathrm{d}x},
\label{eq9}
\end{equation}
whereas the angular distribution of the background intensity is given by equation (\ref{eq7}).

\begin{figure*}
\includegraphics[width=95mm]{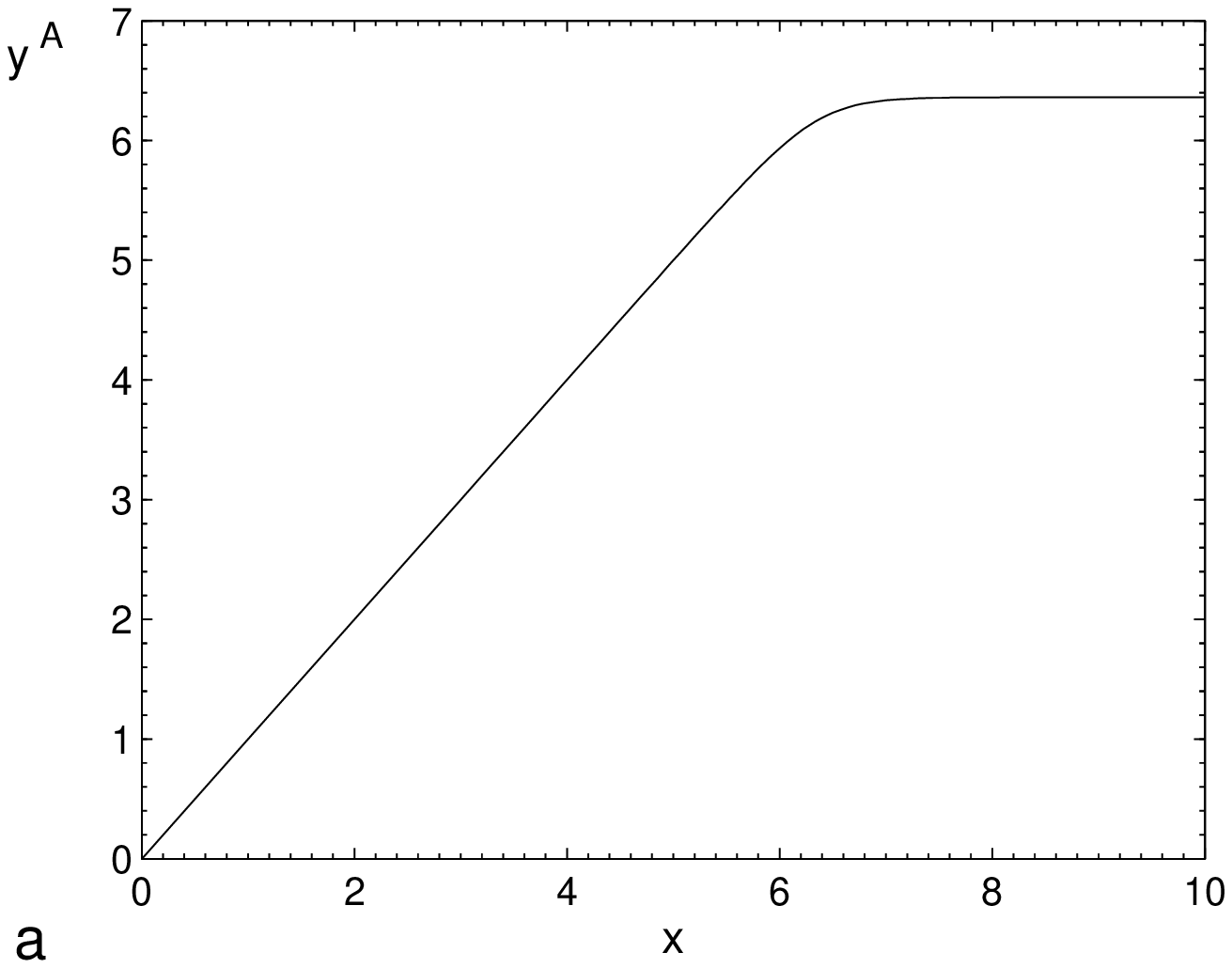}\includegraphics[width=95mm]{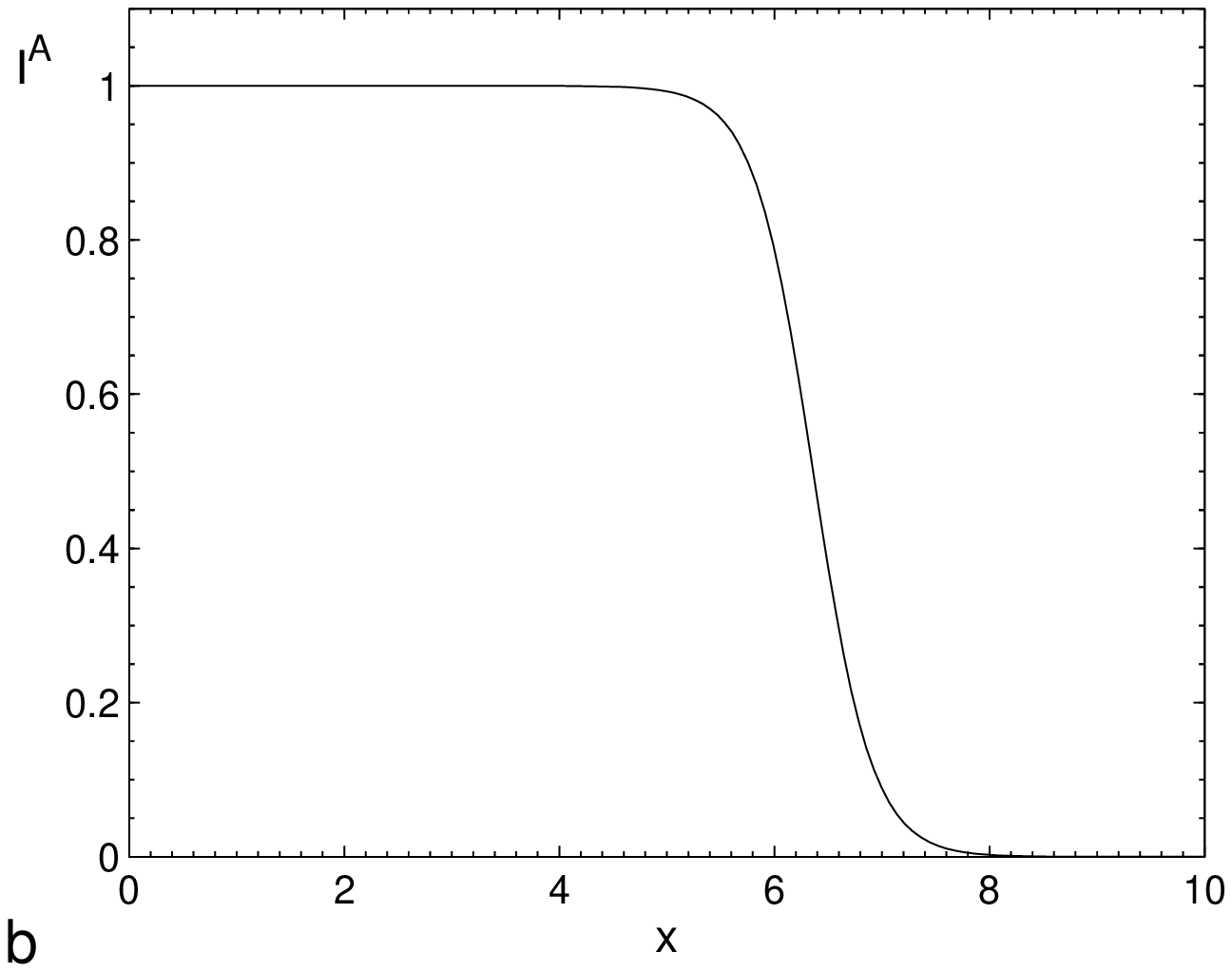}
\includegraphics[width=95mm]{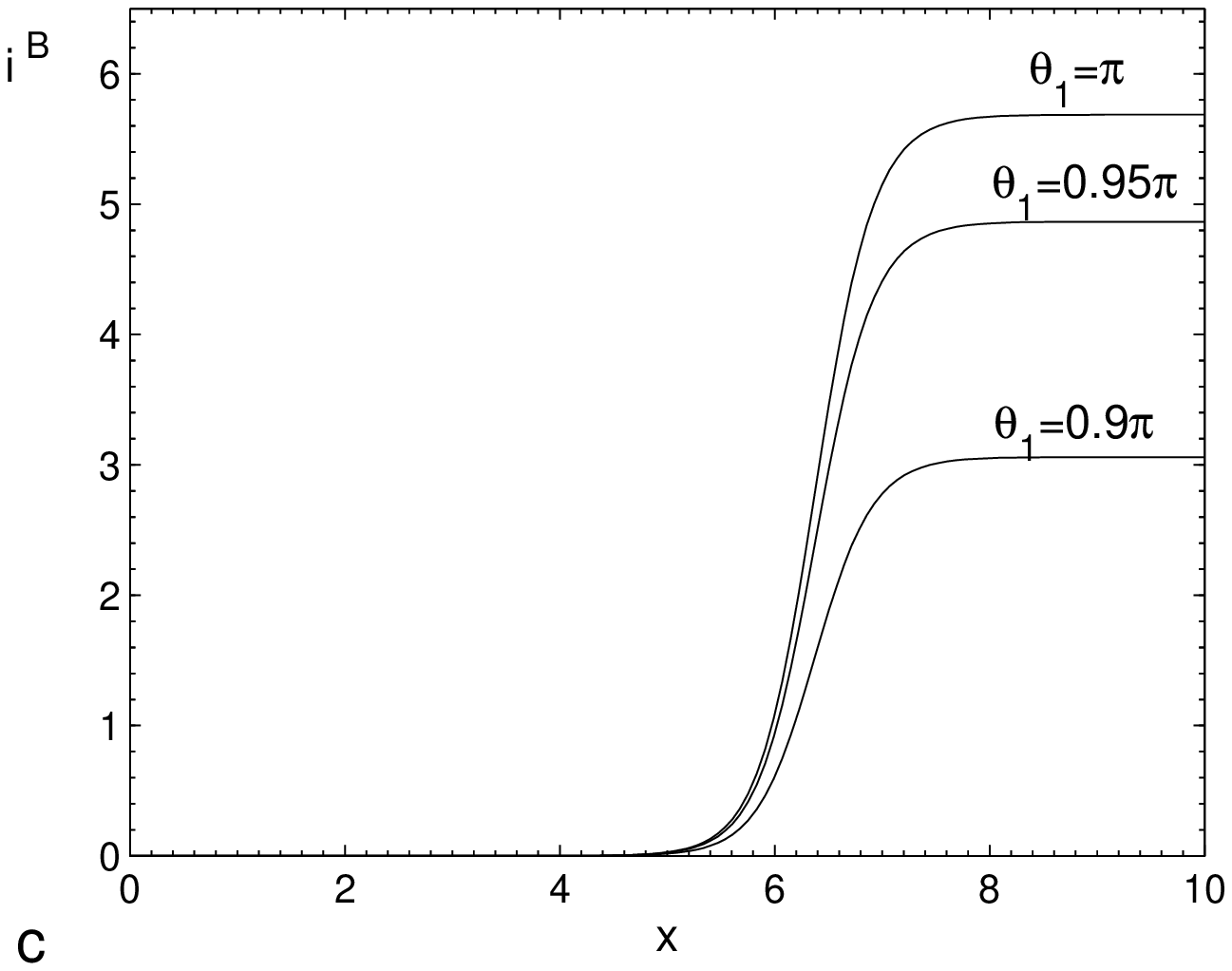}\includegraphics[width=95mm]{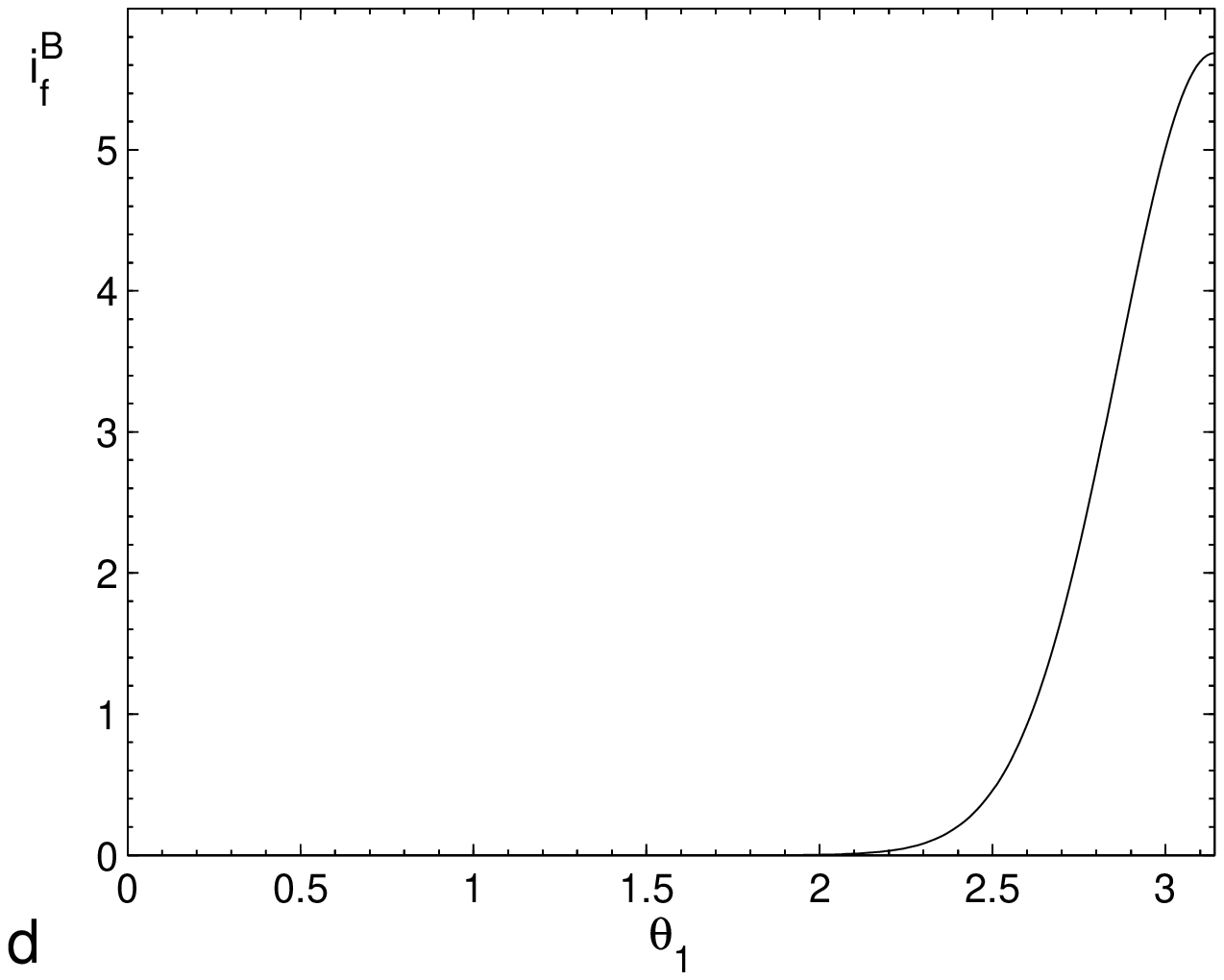}
\caption{Intensity transfer as a result of induced scattering: {\bf a} -- numerical solution of the differential equation (\ref{eq8}); {\bf b} -- beam intensity in the A-polarization as a function of scattering efficiency; {\bf c} -- angular intensity of the background in the B-polarization vs. scattering efficiency for different directions $\theta_1$ ($\Delta\phi=0$); {\bf d} -- final angular intensity of the scattered radiation in the B-polarization vs. photon direction ($\Delta\phi=0$); $\gamma=100$, $I_{\mathrm{bg}}/I_0=10^{-10}$ and $\theta=0.3$.}
\label{f1}
\end{figure*}

The differential equation (\ref{eq8}) does not allow the analytical treatment. The numerical solution is plotted in Fig.~\ref{f1}a. Here and hereafter in this section we concentrate on the case of the A-polarized beam, keeping in mind that for any of the two polarizations the results are qualitatively similar and their quantitative difference is insignificant. The intensity evolution of the beam and the background is shown in Figs.~\ref{f1}b and \ref{f1}c, respectively. One can see that the intensity transfer becomes significant at the scattering efficiency $x\la 10$. Note that for different directions $\theta_1$ the background intensities arrive at the stage of saturation simultaneously, the final intensity values being substantially different.

The difference of the intensity growth in the two polarizations is introduced by the factor $g^{ij}$ (see equation (\ref{eq7})). As can be noticed from equation (\ref{eq4}), for $\theta_1\sim\pi$ the ratio $g^{AA}/g^{AB}$ depends on $\theta_1$ very weakly, but varies with $\Delta\phi$ within a factor of 4, which appears crucial for the relative intensity growth. Given that $\Delta\phi$ changes by $90^\circ$, the value of $g^{AA}/g^{AB}$ is inverted. Thus, the dominant polarization of the scattered radiation is determined by $\Delta\phi$. In Fig.~\ref{f1}c, $\Delta\phi=0$ is taken and the B-polarized intensities are shown, the A-polarized intensities being negligible. The final intensity of the background as a function of $\theta_1$ is presented in Fig.~\ref{f1}d. One can see that the angular distribution of the scattered intensity has a small but meaning width.

\section{Application to the Crab pulsar}
\protect\label{s3}

To get a notion about the efficiency of intensity transfer as a result of the transverse induced scattering in the Crab pulsar one can use the numerical estimate of the quantity $x(\equiv\Gamma)=I_\nu^{(0)}ar$, which is given by equation (12) in \citet{p08b}. Substituting the parameters of the Crab pulsar (the period $P=0.033$ s, the magnetic field strength at the surface of the neutron star $B_\star=4\times 10^{12}$ G, the radio luminosity at 400 MHz $L_{400}=1.4\times 10^{29}$ erg$\,$s$^{-1}$ and the spectral index $\alpha=3$), taking the multiplicity factor of the plasma $\kappa=10^4$ \citep{hm02} and the characteristic Lorentz-factor of the scattering particles $\gamma=10$ and assuming the angle of incidence of the radio emission $\theta=0.3$, we find that at $\nu=1400$ MHz $x=35$. Although the characteristics of the pulsar plasma are very uncertain, one can conclude that in the Crab pulsar the scattering can indeed be efficient.

The scattering efficiency depends on the frequency of the incident radiation. As can be seen from equation (\ref{eq3}), $a\propto\nu^2$, and, taking into account the power-law spectrum of pulsar radiation, $I_\nu^{(0)}\propto\nu^{-\alpha}$, we have $x\propto\nu^{2-\alpha}$. Hence, if we are interested in the scattered component of a given frequency $\nu_1$ ($\nu_1\equiv \nu\eta/\eta_1$), the scattering efficiency $x\propto\eta_1^{2-\alpha}$. Using this in equation (\ref{eq7}), one can obtain the angular distribution of the scattered intensity at the frequency $\nu_1$. 

It should be kept in mind, however, that the formalism of Sect.~\ref{s2} is written in the frame corotating with the neutron star at the angular velocity $\bmath{\Omega}$. To analyse the observational consequences of the induced scattering it is necessary to perform the transformation to the laboratory frame. As a result of this transformation, the corrections to all the quantities involved, except for the photon orientations, are of the second and higher orders in $r/r_L$ (here $r_L$ is the light cylinder radius) and can be ignored in our consideration. The effects of rotational aberration and retardation introduce the first-order corrections and both should be taken into account. Note that the aberration can cause not only quantitative but also qualitative changes in the profiles of the scattered components seen by an observer, and therefore this effect is of an especial interest.

\subsection{The effect of rotational aberration \protect\label{s3.1}}

Generally, the rotational aberration acts to redistribute the intensity, shifting and concentrating the rays toward the direction of the rotational velocity $\bmath{\Omega}\times\bmath{r}$. Note that the rays inclined at the angles $180^\circ\pm\zeta$ to $\bmath{\Omega}\times\bmath{r}$ are shifted in opposite directions, giving rise to the two humps on the intensity distribution. The effects of rotational aberration and retardation in pulsar magnetosphere have been addressed in a number of papers \citep[see, e.g.,][]{m83,kd83,ha01a,gg01,drh04,w06}. Here we proceed from the formalism developed in \citet{p08b} for the rays backscattered along the magnetic lines, generalize that analysis for the rays of an arbitrary orientation and examine the influence of the rotational aberration on the intensity distribution.

\begin{figure*}
\includegraphics[width=130mm]{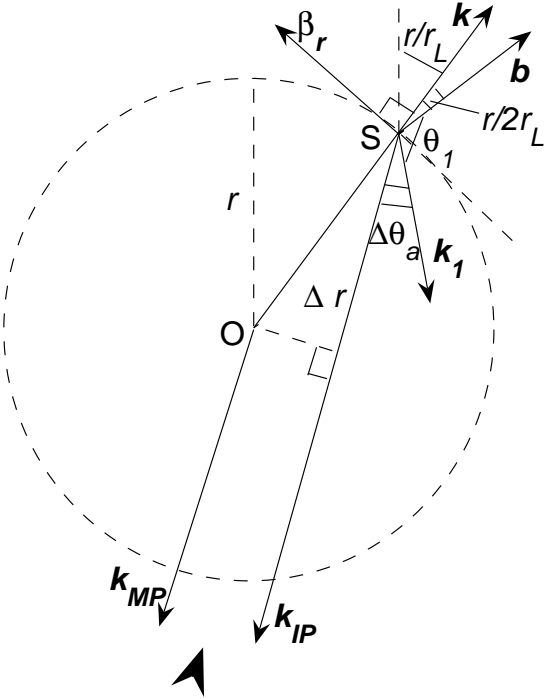}
\caption{Geometry of transverse scattering in pulsar magnetosphere. The pulsar is assumed to be an orthogonal rotator with the rotational axis perpendicular to the plane of the figure and the magnetic axis rotating counterclockwise. For more details see the text.}
\label{f2}
\end{figure*}

As the regime of transverse scattering holds in the outer magnetosphere, one can assume that the emission altitude is negligible as compared to the altitude of the scattering region and the ray geometry is dominated by the effects of magnetosphere rotation, which include the rotational aberration in the scattering region and the retardation because of ray propagation to the scattering point in the rotating magnetic field. For the sake of simplicity we take that the magnetic and rotational axes of the pulsar are perpendicular. The geometric scheme of the scattering is shown in Fig.~\ref{f2}. The ray $\bmath{k}$ emitted along the magnetic axis at $t=0$ at the point O comes to the point of scattering S at $t=r/c$, while the magnetic axis turns by the angle $\Omega r/c\equiv r/r_L$. The polar angle of the point of scattering with respect to the instantaneous magnetic axis is $r/r_L$, and, in the dipolar geometry, the ambient magnetic field vector $\bmath{b}$ is inclined to the axis at the angle $3r/2r_L$. Then the angle between $\bmath{k}$ and $\bmath{b}$ equals $r/2r_L$. At $t=r/c$, the radiation is emitted at point O along the instantaneous magnetic axis and the scattered radiation arises at point S at the angle $3r/2r_L+\theta_1+\Delta\theta_a$ to that direction (here $\Delta\theta_a$ is the correction for the rotational aberration), so that the scattered radiation lags the original one by $3r/2r_L+\theta_1+\Delta\theta_a$ in rotational phase. It should be noted that the scattered component travels somewhat larger distance to the observer than the MP, $\Delta r =r\cos(\pi-\theta_1-\Delta\theta_a-r/2r_L)$. Hence, the pulse longitude of the scattered component with respect to that of the MP reads
\begin{equation}
\lambda_{\mathrm{sc}}-\lambda_{\mathrm{MP}}=\frac{3}{2}\frac{r}{r_L}+\theta_1+\Delta\theta_a-\frac{r}{r_L}\cos(\theta_1+\Delta\theta_a).
\label{eq10}
\end{equation}
Here we have retained only the quantities of the first order in $r/r_L$, since the higher-order effects are ignored in our consideration.

The relativistic aberration is described by the well-known formula
\begin{equation}
\cos\xi_l=\frac{\cos\xi_c+\beta_r}{1+\beta_r\cos\xi_c},
\label{eq11}
\end{equation}
where $\xi_l$ and $\xi_c$ are the wavevector tilts to the rotational velocity in the laboratory and corotating frames, respectively, and $\beta_r$ is the rotational velocity in units of $c$, $\beta_r=\vert\bmath{\Omega}\times\bmath{r}\vert/c$. Taking into account that
\begin{equation}
\xi_c=\left\{
\begin{array}{cc}
\frac{\pi}{2}+\frac{r}{2r_L}+\theta_1,& \theta_1\leq\frac{\pi}{2}-\frac{r}{2r_L},\\
\frac{3\pi}{2}-\frac{r}{2r_L}-\theta_1,& \theta_1\geq\frac{\pi}{2}-\frac{r}{2r_L}
\end{array}
\right.
\label{eq12}
\end{equation}
and using equation (\ref{eq11}), one can find the correction for the aberration $\Delta\theta_a\equiv\theta_1(\xi_l)-\theta_1(\xi_c)$.

\begin{figure*}
\includegraphics[width=130mm]{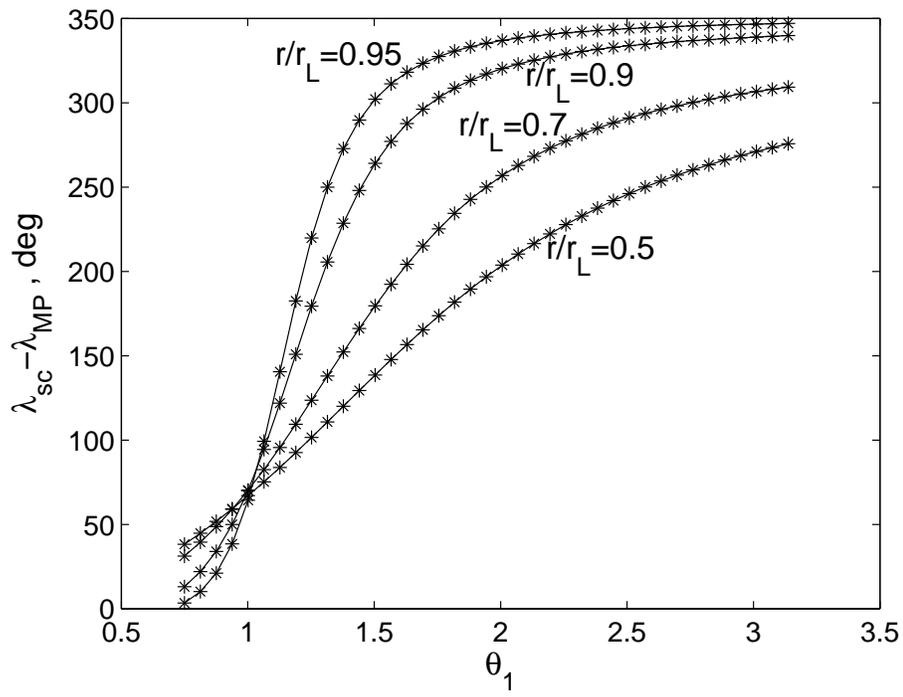}
\caption{Longitudinal location of the rays of equidistant orientations $\theta_1$ on the pulse profile with account for the rotational aberration at different altitudes in the magnetosphere}
\label{f3}
\end{figure*}

The results of numerical calculations based on equations (\ref{eq10})-(\ref{eq12}) are presented in Fig.~\ref{f3}. It shows the longitudinal location of the rays on the pulse profile versus their tilt $\theta_1$ to the magnetic field $\bmath{b}$ for different altitudes of the scattering region. One can see that the rays equidistant in $\theta_1$ tend to concentrate in the two longitudinal regions of the profile, the effect being pronounced for large enough altitudes of the scattering region, in which case the rotational aberration is sufficiently strong. This trend may affect the resultant intensity distribution. The concentration of rays in the region closer to the MP may account for the emission bridge linking the MP and IP. Given that the second region of ray concentration does not coincide with the peak of the original intensity distribution, it may be responsible for a separate component in the pulse profile.

The geometrical considerations leading to equation (\ref{eq10}) concern the transverse scattering of the MP into the IP. However they can also be applicable to the backscattering of the PR and LFC, which are by themselves the results of the MP scattering in the longitudinal regime at lower altitudes in the magnetosphere. It can be shown that if the PR and LFC are subject to the transverse scattering far enough from their origin, $r\gg r_0$, the scattering geometry is deterimned by the magnetosphere rotation rather than by the details of the previous scattering and the longitudinal separations of the resultant backscattered components from the PR/LFC locations on the pulse profile ($\lambda_\mathrm{IP^\prime}-\lambda_\mathrm{PR}$ and $\lambda_\mathrm{HFC}-\lambda_\mathrm{LFC}$) are still given by equation (\ref{eq10}).

\subsection{Profiles of the backscattered components \protect\label{s3.2}}

\begin{figure*}
\includegraphics[width=95mm]{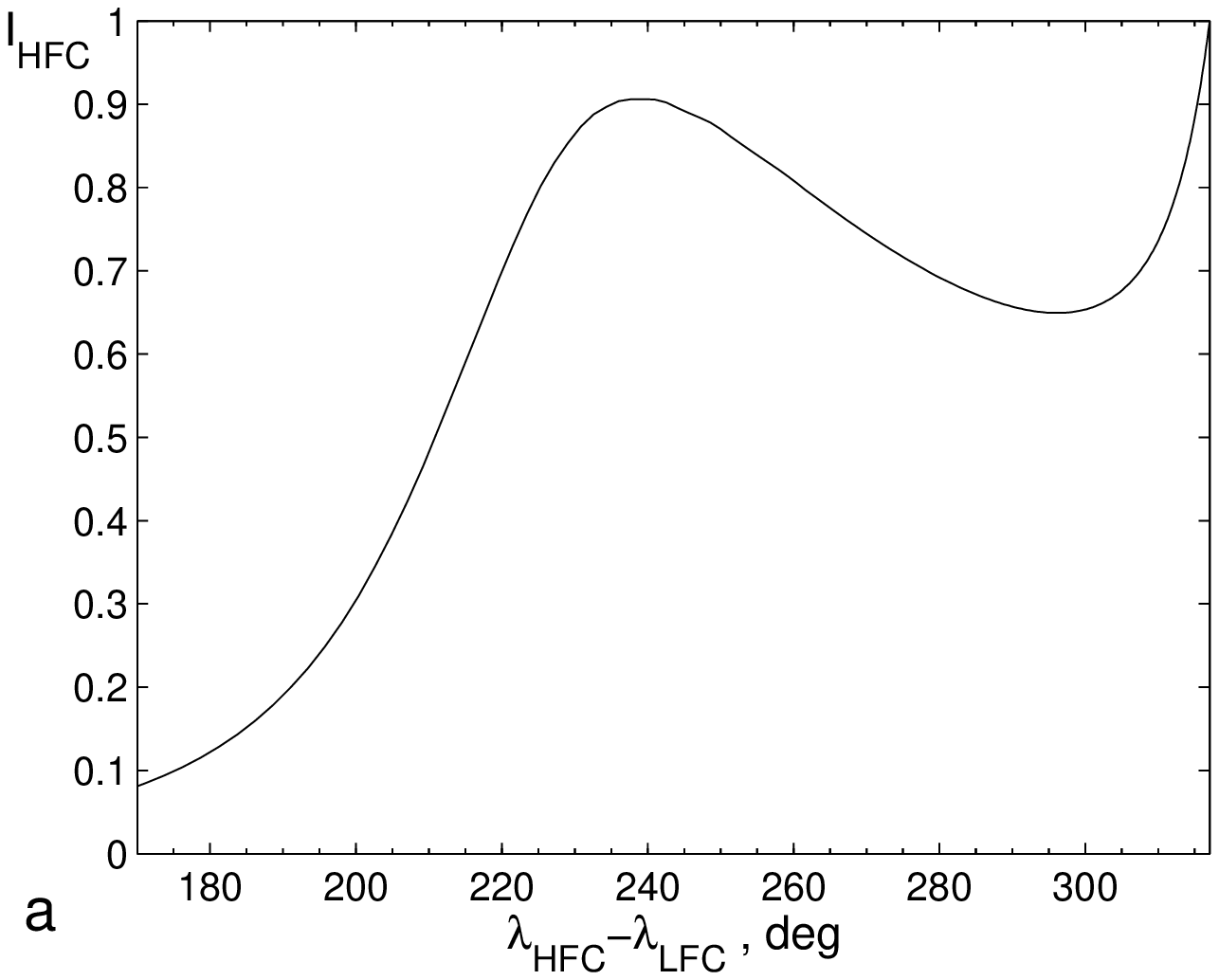}
\includegraphics[width=95mm]{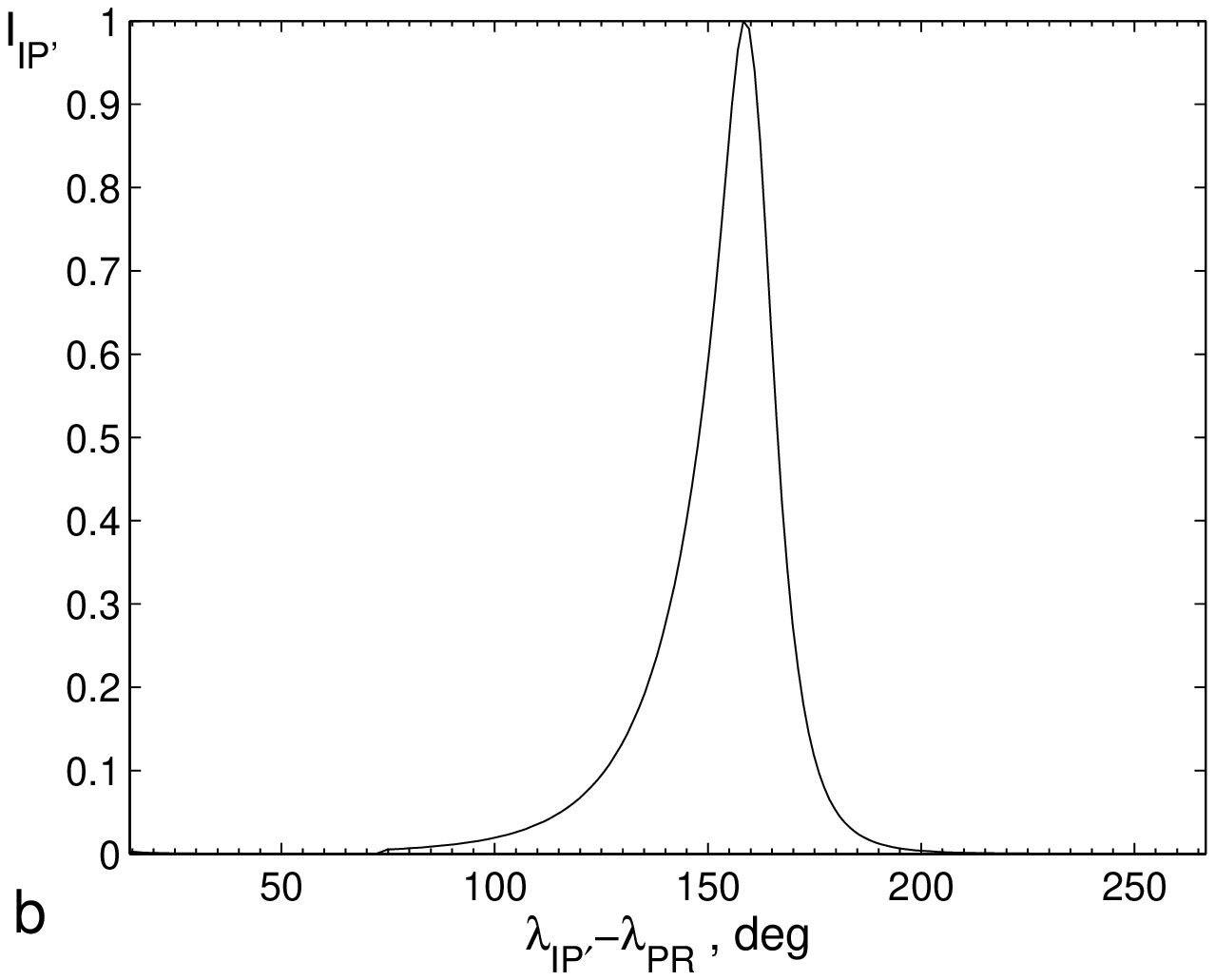}
\includegraphics[width=95mm]{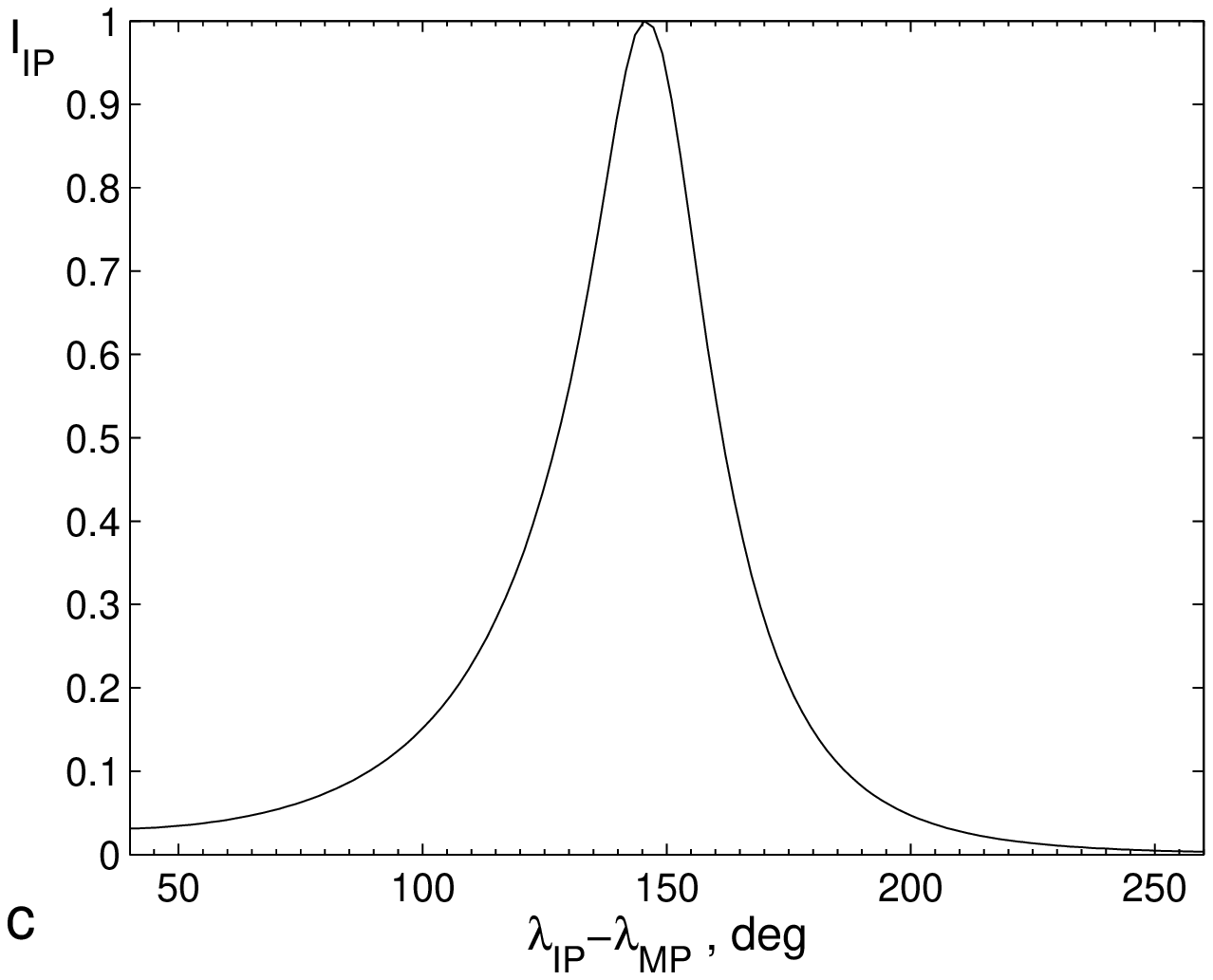}
\caption{Simulated profiles of backscattered components: {\bf a} -- the HFC1 and HFC2 (the result of the scattering LFC$^A\to$HFC$^B$, where the superscripts mark the polarization type, the spectrum of the incident component is $\alpha_\mathrm{LFC}=-3.15$, the scattering efficiency at $\theta_1=\pi/2$ is $x_c=7$, $\Delta\phi=0$, $r/r_L=0.75$); {\bf b} -- the IP$^\prime$ (the scattering PR$^A\to$IP$^{\prime^B}$, $\alpha_\mathrm{PR}=-5$, $x_c=10$, $\Delta\phi=0$, $r/r_L=0.45$); {\bf c} -- the IP (the scattering MP$^B\to$IP$^A$, $\alpha_\mathrm{MP}=-4$, $x_c=5$, $\Delta\phi=0$, $r/r_L=0.6$).}
\label{f4}
\end{figure*}

The numerically simulated profiles of the backscattered components of the Crab pulsar are shown in Fig.~\ref{f4}. The location and shape of the components are determined by the 3 parameters: the scattering efficiency $x_c$ for a fixed $\theta_1$, the spectral index $\alpha$ of the incident radiation and the altitude of the scattering region, $r/r_L$. The scattering efficiency affects the peak location, since it determines the orientations of the rays arriving at the stage of saturation. Larger $x_c$ shift the scattered component toward later pulse longitudes. The steeper the spectral index of the incident radiation, the narrower is the scattered component. The quantity $r/r_L $ determines the strength of the effect of rotational aberration, as is discussed above.

As the LFC originates high in the magnetosphere \citep{p08c}, its scattering to the HFC takes place close enough to the light cylinder, and the rotational aberration plays a significant role. As can be seen in Fig.~\ref{f4}a, it shifts the scattered component as a whole to later pulse longitudes (cf. Fig.~\ref{f3}) and gives rise to an additional peak (known as HFC2) in the intensity distribution. The PR is backscattered into the IP$^\prime$ at lower altitudes, where the rotational aberration is much less significant. The IP$^\prime$ peaks at earlier pulse longitudes (see Fig.~\ref{f4}b) and, due to the extremely steep spectrum of the PR, $\alpha_{\mathrm{PR}}=5$, looks narrow. The scattering of the MP into the IP is qualitatively similar, but the MP needs much steeper spectrum than is observed to provide the IP narrowness. It should be noted that the induced scattering may flatten the spectrum of the incident radiation, being more efficient at lower frequencies. Hence, the original spectrum of the MP may really be steeper. In our simulation presented in Fig.~\ref{f4}c we have assumed $\alpha_{\mathrm{MP}}=4$, but it still seems smaller that necessary.

Note that for the scatterings PR$\to$IP$^\prime$ and LFC$\to$HFCs we assume the A-polarization of the incident radiation, in which case the scattered radiation is dominated by the B-polarization. The polarization type of the PR and LFC is unambiguously determined by the physics of longitudinal scattering which gives rise to these components (see Sect.~\ref{s2.1}). The position angle shifts of the IP$^\prime$ and HFCs by $90^\circ$ from those of their feeding components is in line with the observations \citep{mh99}.

\section{Discussion}
\protect\label{s4}

\begin{figure*}
\includegraphics[width=130mm]{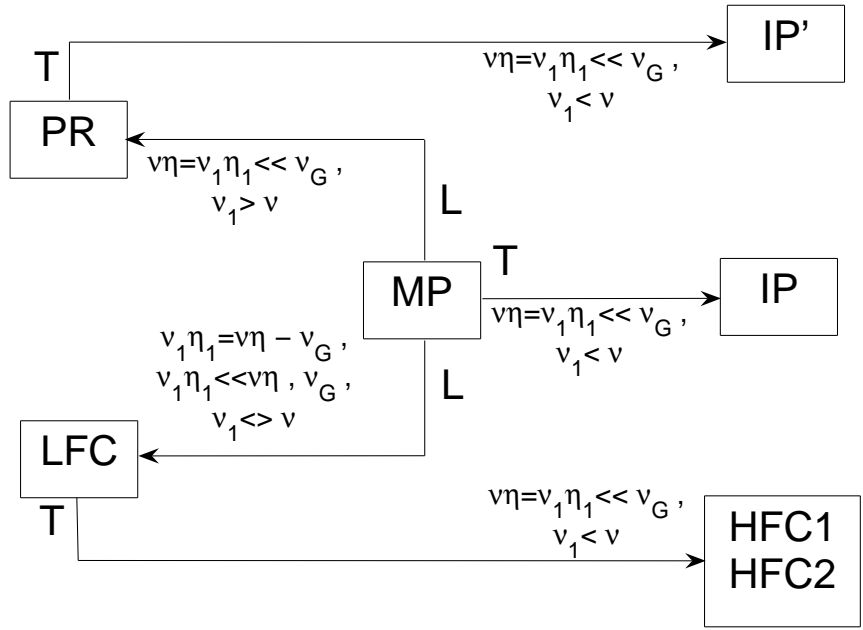}
\caption{Summary of the component formation in the Crab pulsar. The marks L and T refer to the longitudinal and transverse regimes of induced scattering, respectively.}
\label{f5}
\protect\label{lastpage}
\end{figure*}

We have suggested the scenario of the component formation in the Crab pulsar (see Fig.~\ref{f5} for summary). It is based on the induced scattering of the MP into the background, which takes place in the flow of the secondary plasma inside the open field line tube of the pulsar. Deep in the magnetosphere, the magnetic field is so strong that the MP radiation is scattered in the longitudinal regime and gives rise to the component directed approximately along the streaming velocity of the scattering particles. It appears on the pulse profile ahead of the MP and can be identified with the PR \citep{p08a,p08c}. As the magnetic field strength decreases with distance, at higher altitudes the scattering switches to the transverse regime and results in the component roughly antiparallel to the particle velocity, which can be recognized as the IP \citep{p08b}. Resonant absorption of the radio emission leads to the relativistic gyration of the scattering particles, and the scattering from the first harmonic of the gyrofrequency to the zeroth one can account for the LFC \citep{p08c}. On the way in the magnetosphere, the PR and LFC are subject to the transverse scattering, which gives rise to the IP$^\prime$ and HFCs, respectively.

According to equation (\ref{eq1}), the scattered radiation of a given frequency and different orientations is fed by the incident radiation of substantially different frequencies. The scattering efficiency $x$ depends on the incident intensity, and hence, the steep spectrum implies an essential distinction of $x$ at different pulse longitudes of the scattered component. It is the effect that proved to determine the peak locations of the backscattered components of the Crab pulsar (see Sect.~\ref{s3}). In case of the MP scattering to the PR and LFC, however, the scattering efficiency appears much stronger function of the ray orientation, and the components always peak at $\theta_1\sim 1/\gamma$, independently of the spectrum of the incident radiation.

The observed distribution of the scattered intensity can be noticeably affected by the rotational aberration, which typically shifts and concentrates the rays toward later pulse longitudes. The effect is especially pronounced in the HFCs, which are believed to originate close to the light cylinder. In the framework of our model, they present a single component, with one of the peaks (the HFC1) being determined by the efficiency of induced scattering and another one by the rotational aberration. In the case of the MP scattering to the leading components (the PR and LFC), at a fixed altitude the scattered intensity has almost delta-functional angular distribution. Then the role of rotational aberration is to place these components ahead of the MP and to provide their finite widths by shifting the radiation originating at somewhat different altitudes to different pulse longitudes.

\subsection{Frequency evolution of the profile \protect\label{s4.1}}

The profile of the Crab pulsar drastically changes with frequency \citep[e.g.,][]{mh96}: each of the components is visible only over the part of the radio frequency band. In our model, the frequency evolution of the components can be understood as follows. As the characteristic altitudes of the scattering regions are related to the cyclotron resonance radius, lower frequencies are scattered somewhat higher in the magnetosphere. The scattering MP$\to$LFC holds at $r\approx r_L$, and for the lowest frequencies the scattering site is expected to lie beyond the light cylinder, in which case the stable component is not formed. Therefore the LFC appears on the pulse profile at high enough frequencies. The re-scattering of the LFC into the HFCs takes place inside the light cylinder only for still higher frequencies. Hence, at the highest radio frequencies the HFCs become prominent, whereas the LFC vanishes.

The efficiency of the transverse scattering explicitly scales with frequency as $x\propto\nu^{2-\alpha}$, and since $\alpha_{\mathrm{MP}}=3$, the scattering MP$\to$IP is believed to be significant only at low enough frequencies. At higher frequencies it is dominated by the scattering MP$\to$PR$\to$IP$^\prime$. It should be pointed out that the latter process takes place at lower altitudes. Indeed, the longitudinal scattering (MP$\to$PR) occurs lower than the transverse one (MP$\to$IP), since the magnetic field strength decreases with distance. The frequency of the PR is much higher than that of the MP, $\nu_\mathrm{PR}\sim\nu_\mathrm{MP}\theta_\mathrm{MP}^2/\theta_\mathrm{PR}^2\sim\nu_\mathrm{MP}\theta_\mathrm{MP}^2\gamma^2$ (see equation (\ref{eq1})), and the PR tilt to the ambient magnetic field, $\theta_\mathrm{PR}$, rapidly increases along the trajectory from $\sim 1/\gamma$ to $\sim r/r_L$, so that this component meets the condition of transverse scattering very soon. Thus, the IP$^\prime$ originates at lower altitudes than the IP, the former process being more efficient, in particular, because of larger number density of the scattering particles, $n_e\propto r^{-3}$. Probably, it is the altitudinal dependence of the scattering efficiency that provides stronger transverse scattering at higher frequencies (i.e. at lower altitudes) and accounts for the increasing spectrum of the IP$^\prime$ and the flat spectrum of the HFCs.

Note that the spectra of the components subject to the scattering should be substantially altered by this process, so that the original spectra of the MP, PR and LFC differ from the observed ones and remain obscure. The spectral indices used in our simulations (see Fig.~\ref{f4}) are taken only for illustrative purposes. A detailed analysis of the spectra of the backscattered components is also beyond the framework of the present paper.

The observations have revealed several fine effects in the high-frequency profiles: as the frequency increases, the IP$^\prime$ widens, the HFCs shift to later pulse longitudes, and the HFC2 becomes somewhat stronger than the HFC1 \citep{mh99}. In principle, all these effects can be reproduced by means of a slight variation of the parameters taken to construct the profiles in Fig.~\ref{f4}. However it should be kept in mind that our consideration is too simplified, since we have assumed a fixed altitude of the formation of a backscattered component and neglected the width of an incident component as compared to its tilt to the ambient magnetic field. The fine effects should be considered in a more accurate model.

\subsection{Polarization properties and geometry of the pulsar \protect\label{s4.2}}

The position angle swing across the profile of the Crab pulsar cannot be explained in terms of the customary rotating vector model \citep{mh99,kjw04}. Evidently, the assumption that all the profile components originate at the same altitude is not valid in this case. In the framework of our model of the Crab emission pattern, the profile components arise at different sites in the magnetosphere, the radiation  can be directed antiparallel to the particle velocity, and the scattering can introduce a $90^\circ$-shift of the position angle. Then the polarization picture roughly looks as follows.

Very little change of the position angle across the MP testifies to the nearly central cut of the emission cone by the sight line. In the PR and IP components, which result from the MP scattering, the position angle is almost the same. The position angle of a scattered component reflects that of the incident radiation in the scattering region. Although the PR and IP originate at high enough altitudes, the position angle swing is still determined by the very small impact angle of the observer to the magnetic axis, $\beta$, and is not modified substantially (i.e. $(r/r_L)^2\sin\alpha\cos\alpha\ll\beta$, where $\alpha$ is the angle between the magnetic and rotation axes of the pulsar).

It should be noted that, in accordance with the physics of longitudinal scattering, the PR component should be characterized by complete linear polarization of the A-type. The ordinary polarization of the PR, along with the MP and IP, is confirmed by the high-energy observations of the Crab PWN \citep*[e.g.,][]{lai01}. The transverse scattering of the PR leads to the IP$^\prime$ with almost complete B-polarization (cf. equation (\ref{eq4})). Apart from the $90^\circ$-shift, the position angle of the IP$^\prime$ is believed to roughly reflect that of the PR.

The LFC shows an essentially distinct behaviour of the position angle: it increases linearly with pulse longitude, and the total shift from the MP position angle is $\Delta\psi\approx\lambda_\mathrm{MP}-\lambda_\mathrm{PR}\approx 30^\circ$. This can be presumably attributed to the peculiar location of the LFC origin -- close to the light cylinder. For an arbitrary angle $\alpha$ between the magnetic and rotational axes (given that $(r/r_L)\sin\alpha\cos\alpha\gg\beta$) far from the emission region the rays make the angle $(r/2r_L)\sin\alpha$ with the ambient magnetic field, and the position angle is $\Delta\psi=(r/2r_L)\cos\alpha$. Then the component scattered along the ambient magnetic field appears on the pulse profile at the longitude $\lambda_\mathrm{LFC}=\lambda_\mathrm{MP}-(r/2r_L)\sin\alpha$ and has the same position angle, $\Delta\psi=(r/2r_L)\cos\alpha$. (Let us note in passing that in an arbitrary geometry of the pulsar the backscattered components may also be visible provided that their orientation is determined by the pulsar spectrum rather than by the ambient magnetic field direction, as is suggested in the present paper.) Thus, the linear increase of the position angle with pulse longitude is indeed the case, $\Delta\psi\propto\lambda_\mathrm{MP}-\lambda_\mathrm{LFC}\propto r/r_L$, but $\Delta\psi\approx\Delta\lambda\approx 30^\circ$ cannot be achieved for any $\alpha$. The value $\alpha\approx 60^\circ$ cited in the literature \citep[e.g.,][]{mh99} yields only rough correspondence. In the more plausible case, $\alpha\approx 90^\circ$, we have $\Delta\psi\to 0$. The observed linear increase of $\Delta\psi$ with $\Delta\lambda\propto r/r_L$ is suggestive of another first-order effects, such as the action of the current flow \citep{ha01a}, though the higher-order effects close to the light cylinder cannot be excluded as well.

The position angle of the HFCs is shifted by $\sim 90^\circ$ from that of the LFC, so that the A-polarized LFC is chiefly scattered into the B-polarization. In the two components, the HFC1 and HFC2, the position angle changes within the same range, confirming their origin as a single component affected by the rotational aberration. More observational details given in \citet{mh99}, e.g. different slopes of the position angle swing in the HFCs, cannot be compared to our model because of its restrictions (recall that we have assumed a fixed altitude of the component formation and neglected the width of the component subject to scattering).

\subsection{Giant pulses \protect\label{s4.3}}

Similarly to the MP, all the backscattered components of the Crab pulsar -- the IP, IP$^\prime$ and HFCs -- exhibit occasional giant pulses, whereas the PR and LFC do not \citep{c04,j05}. In the framework of our model, this can be understood as follows. As the efficiency of induced scattering is determined by the incident intensity, giant MPs are scattered much stronger than the normal ones, and a substantial part of their energy may be transmitted to the scattered components, giving rise to the giant IPs, PRs and LFCs. Later on the giant PRs and LFCs are re-scattered into the IP$^\prime$s and HFCs, depositing most of their energy to these components and causing the giant IP$^\prime$s and HFCs. Thus, the scattering process transmits the energies of the giant MPs between different components, and these energies are never stored in the intermediate components (the PR and LFC). It should be kept in mind that the energy of giant MPs is transferred between widely spaced frequencies (cf. equation (\ref{eq1})), and, given that the giant MPs are sufficiently narrowband, at a given frequency the giant pulses in different profile components are not simultaneous.

The induced scattering modifies the temporal and frequency structure of the incident radiation \citep[see also][]{p08b}. As is found in \citet{eh07}, the giant MPs and IP$^\prime$s of the Crab pulsar actually exhibit quite distinct structure. The giant MPs present the broadband microbursts built of narrowband ($\delta\nu/\nu\sim 0.1$) nanoshots of a duration $\delta t\sim 10/\nu\sim 10^{-8}-10^{-9}$s. The giant IP$^\prime$s consist of the proportionally spaced narrow emission bands of microsecond lengths, which drift noticeably toward higher frequencies. In our model, the distinction between the giant MPs and IP$^\prime$s can be interpreted in terms of equation (\ref{eq1}), which relates the frequency and orientation of the scattered radiation to those of the incident radiation. For the scattering MP$\to$PR$\to$IP$^\prime$ it takes the form
\begin{equation}
\nu_\mathrm{MP}\theta^2/2=\nu_\mathrm{PR}/\gamma^2=\nu_\mathrm{IP^\prime}(1-\beta\cos\theta_1).
\label{eq13}
\end{equation}
Differentiation of the extreme terms of equation (\ref{eq13}) with respect to frequency yields $\Delta\nu_\mathrm{MP}/\nu_\mathrm{MP}=\Delta\nu_\mathrm{IP^\prime}/\nu_\mathrm{IP^\prime}$. Thus, a nanoshot with $\delta\nu/\nu\sim 0.1$ is scattered into the same normalized band $\delta\nu/\nu\sim 0.1$, and the scattered radiation strongly concentrates toward the edge of the band because of the frequency dependence of the scattering efficiency. Then the groups of the nanoshots in the MP create the sets of emission bands in the IP$^\prime$.

The transformation of the angular scale of a nanoshot as a result of the scattering can be found by differentiating equation (\ref{eq13}) with respect to angle:
$2\Delta\theta/\theta=\beta\sin\theta_1\Delta\theta_1/(1-\beta\cos\theta_1)$. Given that $\pi-\theta_1\ll 1$, this yields $\Delta\theta_1\approx 4\Delta\theta/\theta(\pi-\theta_1)\gg\Delta\theta$. Thus, the angular scale of the scattered radiation increases by about two orders of magnitude. Keeping in mind that the radiation scattered from different nanoshots may overlap, one can expect that the group of nanoshots in the MP gives rise to the feature of microsecond length in the IP$^\prime$.

As can be seen from equation (\ref{eq13}), the radiation with somewhat different angles of incidence $\theta$ (i.e. different parts of a nanoshot or the neighbouring nanoshots) is scattered to somewhat different frequencies $\nu_\mathrm{IP^\prime}$, leading to the upward drift of the emission bands in the IP$^\prime$. Thus, the peculiar temporal and frequency structure of the IP$^\prime$ can be attributed to the induced scattering of the MP.

\section{Conclusions}
\protect\label{s5}

We have examined the induced scattering of a bright narrow radio beam into a wide solid angle. The radiation is scattered off the particles of the ultrarelativistic electron-positron plasma streaming along the open field lines in the magnetosphere of a pulsar. Our consideration is concerned with the transverse scattering in the presence of a moderately strong magnetic field. The intensity of the radio beam of a given frequency is transferred to the background radiation of the frequencies and orientations related by equation (\ref{eq1}), and the scattered intensity peaks in the direction antiparallel to the velocity of the scattering particles. The background radiation of a given frequency and different orientations $\theta_1$ is fed by the beam radiation of different frequencies, and therefore the efficiency of intensity transfer to different $\theta_1$ is determined not only by the scattering geometry but also by the corresponding original intensity of the radio beam. Hence, the angular distribution of the scattered radiation is affected by the radio beam spectrum, and for steep enough spectra of the pulsar radio emission it is the effect that determines the location of the scattered component on the pulse profile.

Based on the numerical solution of the problem of intensity transfer we have simulated the profiles of the backscattered components of the Crab pulsar. It is suggested that the IP, IP$^\prime$ and HFCs result from the transverse scattering of the MP, PR and LFC, respectively. As the scattering regions lie in the outer magnetosphere, the observed profiles can be noticeably affected by the rotational aberration. A detailed treatment of this effect shows that not only the components are shifted in pulse longitude but also their intensities can be substantially redistributed. In particular, it is argued that the HFCs are a single component, with the second peak (the HFC2) arising as a result of the aberration.

As the altitude of the scattering region is related to the radius of cyclotron resonance, higher frequencies are scattered somewhat lower in the magnetosphere. The number density of the scattering particles changes as $r^{-3}$, and the incident intensity as $r^{-2}$, so that at lower altitudes the scattering is more efficient. The spectral behaviour of the backscattered components is presumably determined by the altitudinal dependence of the scattering efficiency: the IP$^\prime$ and HFCs appear in the high-frequency profiles and have flat or even increasing spectra. Note that at high frequencies the scattering MP$\to$IP is dominated by the scattering MP$\to$PR$\to$IP$^\prime$, which takes place at lower altitudes, whereas the scattering MP$\to$LFC$\to$HFCs occurs inside the light cylinder and forms a stable component.

Our model of the component formation in the Crab pulsar is supported by the polarization data. The PR and the result of its backscattering, the IP$^\prime$, both have complete linear polarization and the position angles differing by $90^\circ$. The position angles of the LFC and HFCs also differ by $90^\circ$. The flat position angle swing across the PR, MP and IP is suggestive of an almost central cut of the pulsar emission cone by the sight line. Given that the Crab pulsar is a nearly orthogonal rotator, the linear increase of the position angle with pulse longitude observed in the LFC as well as the total position angle shift of $\sim 30^\circ$ from that of the MP can be attributed to the effect of the current flow close to the light cylinder. Note that all the scattered components can be observable for an arbitrary (but not very small) angle between the magnetic and rotation axes of the pulsar.

In the framework of our model of the component formation in the Crab pulsar, the giant pulses characteristic of the MP can also be seen in the scattered components, since stronger incident intensities imply higher efficiency of intensity transfer to the scattered components. The lack of giant pulses in the PR and LFC can be understood as a consequence of their re-scattering to the IP$^\prime$ and HFCs. The induced scattering can also account for the observed difference in the frequency and temporal structure of the giant MPs and IP$^\prime$s.


\clearpage

\input epsf

\clearpage

\clearpage

\clearpage

\clearpage

\protect\label{lastpage}

\end{document}